# DESIGNING A LOGICAL SECURITY FRAMEWORK FOR E-COMMERCE SYSTEM BASED ON SOA

Ashish Kr. Luhach[1], Dr. Sanjay K. Dwivedi[2], Dr. C. K. Jha[3]

[1]Dronacharya College of Engineering, Gurgaon, Hr, India
[2]BBA University, Lucknow, U.P., India
[3]Banasthali University, Jaipur, Rajasthan, India.

*Abstract*

*Rapid increases in information technology also changed the existing markets and transformed them into e-markets (e-commerce) from physical markets. Equally with the e-commerce evolution, enterprises have to recover a safer approach for implementing E-commerce and maintaining its logical security. SOA is one of the best techniques to fulfill these requirements. SOA holds the vantage of being easy to use, flexible, and recyclable. With the advantages, SOA is also endowed with ease for message tampering and unauthorized access. This causes the security technology implementation of E-commerce very difficult at other engineering sciences. This paper discusses the importance of using SOA in E-commerce and identifies the flaws in the existing security analysis of E-commerce platforms. On the foundation of identifying defects, this editorial also suggested an implementation design of the logical security framework for SOA supported E-commerce system.*

## 1. INTRODUCTION

The customarily architecture models used for E-commerce modeling are Common Object Request Broker Architecture (CORBA) and Component Object Model (COM). These architectures are based on distributed object technology having different data standards and protocols, which are relatively complex to implement. These architecture models are tightly coupled in nature which means if new business associates have to tie into the existing system or codification, the whole existing code or system has to redevelop to add the young business associate. This redevelopment brings in high operational cost and low performance to the whole organization. These architectures are not compatible with modern E-commerce systems which call for loose coupling, flexibility and other demands of dynamic commercial enterprise applications or environments.

Service Oriented Architecture (SOA) fulfills the demands of modern E-commerce, such as loose coupling of business uses and flexibility of services. SOA enables the infrastructure resources of an endeavor to defend a society of its users and applications through services that are spread athwart the enterprise [1]. SOA can be limited as an approach of planning and building systems that are elastic, adaptable and scalable in order to support dynamic business applications or environments. SOA allows enterprises to deploy, build, design and integrate the services that are independent of applications and platforms on which they operate. These services are linked together through business processes in order to form composite services and applications which execute the line affairs. Each service implements an action or procedure. Services are not affiliated with each other; they are units of functionality which are generally coupled with each other and apply the standard protocols to limit the communication between them. They need the metadata, which identifies the characteristics of all the services as well as the data that drive these services [3].

SOA has advantages of being gentle to use, flexible, reusable, and scalable. These advantages made SOA the paramount choice to implement E-commerce. With the advantages, SOA endows with ease for message tampering and unauthorized access which makes, "the security implementation of E-commerce", more complex and unmanageable. The primary concern of this research is to ascertain the security of SOA based E-commerce. This column is intended to design an implementation; security framework of SOA based E-commerce.

## 2. CURRENT SECURITY STANDARDS AND TECHNOLOGY – RELATED WORK

In this part, we will talk about the various security criteria and technologies for E-commerce system and security attacks related to these technologies. The various security standard and technologies are:

### 2.1 Transport layer security (TLS): Secure Socket Layer (SSL)

Transfer Layer Security (TLS) and its predecessor, Secure Sockets Layer (SSL) is a non XML security framework and widely applied to transport layer data communication. This security standard provides authentication, integrity and confidentiality of the transmitted messages. SSL uses public and secret key cryptography to maintain the safe data communication. SSL can be utilized in different modes according to the security necessities, the popular, modes are No authentication, Server authentication and two way authentications. No authentication means neither the customer nor the server authenticates itself to the other. In such case, only confidentiality is applied. Server authentication means only the server authenticates itself to the customer. Two-way authentication means both client and server authenticate themselves to each other. The various applications of TLS/SSL are E-commerce system (internet sites), Key exchange, Cipher, Web browsers and Libraries

#### 2.1.1 Security Attacks against TLS & SSL

Following are the significant security attacks against TLS/SSL:

- ✓ Renegotiation attacks.
- ✓ Version rollback attacks.
- ✓ BEAST attacks.
- ✓ CRIME and BREACH attacks.
- ✓ Padding attacks.
- ✓ RC4 attacks.
- ✓ Survey of websites.
- ✓ Truncation attacks.

### 2.2 XML Encryption

XML encryption, is a security standard developed by the World Wide Web Consortium (W3C) in 2002. XML encryption is primarily applied for encryption and decoding of XML documents and can used for any kind of information. XML encryption uses various encryption algorithms to code and decode the data elements for example AES, RSA and triple DES are mainly utilized for the same. XML encryption supports both forms of symmetric and asymmetric cryptography. XML encryption ensures the confidentiality of the communicated messages. There are five types of XML encryption, which are Encryption XML Element, Encryption XML Element Content (Elements), Encryption XML Element Content (Character Data), Encryption Arbitrary Data and XML Documents and Super-Encryption.

### 2.2.1 Security Attacks against XML encryption

- ✓ Weak chaining of cipher-text blocks: we can able decrypt the data by sending modified cipher-texts to the server and gather information from the error messages.

## 2.3 XML Signature

As digital signature, XML signatures are applied as a standard for verifying the parentage of the transmitted messages. XML signatures were developed by the W3C with the collaboration of Internet Engineering Task Force (IETF). XML signature can be applied for securing any type of messages, but typically used in an XML document. Any text files which are accessible through Uniform Resource Allocator (URL) can be stopped up by XML signature. XML Signature supports authentication, information integrity, and non-renunciation. The most significant characteristic of an XML signature is it can digitally sign a lot of XML document instead of signing the whole text file as a digital signature. XML signature allows multiple signatures to different portion of an XML document. To validate the XML signature's authenticity a procedure called Core validation is applied. XML signatures are more flexible in comparison with other sorts of signatures as it utilizes the XML content instead of the binary information.

### 2.3.1 Security issues in XML signature

- ✓ XML signature is criticized for their security architecture in general as of their complexity and poor performance.
- ✓ XML signatures are not suitable for sensitive SOA applications.
- ✓ The use of XML signatures in SOAP and WS-security can be vulnerable if not properly implemented.

## 2.4 XML key management specification (XKMS)

XML key management specification (XKMS) is defined as web services, which supply an intermediate interface between XML application and Public Key Infrastructure (PKI). XKMS were developed by the W3C with the collaboration of Internet Engineering Task Force (IETF). XKMS greatly simplifies the deployment of enterprise strength Public Key Infrastructure by transferring complex processing tasks from the client application to a Trust Service [7]. XKMS supports XML signatures and XML encryptions. XKMS basically designed for the distribution and enrollment of public keys. The design criteria of XKMS include the following factors, which are Compatible with XML signature and XML encryption, Implementation of XKMS should be simple as possible using XML tools and minimize the client code and configuration. XKMS was built in two parts, which are XML Key Information Service Specification (X-KISS) are used for processes and validates the public keys and XML Key Registration Service Specification (X-KRSS) used for registration of public keys and provides four services such as revoke, re-issue, register and recover.

### 2.4.1 Security issues in XKMS

- ✓ Reply attacks.
- ✓ Denial of service attacks.
- ✓ Key recovery policy.

### 2.5 Security Assertions Markup Language (SAML)

Security Assertions Markup Language (SAML) is specified as open standard data format, which is based on XML. SAML is used for authentication and authorization of data between communicating parties. SAML developed by the Organization for the Advancement of Structured Information Standards (OASIS) in 2001. Endrei et al. [7] Defines SAML as the first standard widely adopted by industry for securing e-business dealings based on XML. The inspiration behind to develop SAML was to furnish a common means for sharing security services for commercial enterprise to business (b2b) and business to consumer (b2c) transactions. SAML provides better authorization, profile information and certification for the parties involved in the communication.

SAML includes four different components [8], which are: SAML Assertions, it contain the security information such as how authentication and authorization information are defined. It usually transferred to service providers by identity providers. SAML assertion can be utilized to prevent man-in-middle and replay attacks [8]. SAML assertions can be signed by XML signatures and encrypted by XML encryption. It too holds in several statements that used for access control decisions by service providers. The different types of statements used in SAML are Authentication statement, Attribute statement and Authorization decision statement.

SAML Protocols: it defines how SAML requests and response are handled. It also defines various rules for the same. It can be defined as a simple request and response protocol. The most important SAML protocol requests are known as a query. SAML binding it binds the SAML protocol messages with the industry adopted protocols. For example SAML SOAP bind specifies how SAML encapsulated with SOAP bound to SOAP message. SAML Profiles it defines how SAML assertions, protocols and bindings are combined together to support a particular use case.

#### 2.5.1 Security flaws in SAML

SAML recommends a variety of security mechanisms such as:

- ✓ XML signature and XML encryption.
- ✓ TLS and SSL.

### 2.6 Kerberos

Kerberos is used as authentication protocol, which helps to ensure the identity among users communicating over an unsecured network. Kerberos provides mutual authentication in a typical client server environment where customer and server both verifies each other identity. Kerberos provides a secure environment against eavesdropping and reply attacks. Kerberos uses symmetric key cryptography.

#### 2.6.1 Drawbacks of Kerberos

- ✓ Single point of failure.
- ✓ Restricted time requirement.
- ✓ Requires user accounts.
- ✓

## 3. SECURITY ANALYSIS OF SOA BASED E-COMMERCE

The general system architecture of an E - commerce system is presented above in figure 1. E-commerce systems provide users with a variety of business service components and user can call any of these business service components online such as trading information services, contract management services and other information management services, if they are authenticated and passed. Before calling business service component's user submits their authentication through HTTP to a WWW host. Users can log on to the component requested through Simple Object Access Protocol (SOAP) and at the same time user may utilize other services as well, which is furnished by other collaborators such as banking and logistics services. E-commerce platform is hardly used for online transactions, but also can be practiced as a serial publication of services such as banking and revenue enhancement. So, E-commerce provides a complete lay down of business procedures which consist of various inspection and repairs. These services are originated by different vendors and managed or kept up by diverse departments within an endeavor, which makes E-commerce a service oriented heterogeneous system. This heterogeneous system will confront a kind of security threats and risk factors which are:

### 3.1 Certificate Duplicity

Approach to business service components, users should submit their authentication through HTTP and get it validated from web hosts. In return an authentication credential is released for users as a validation of certification. After getting this authentication certificate user can access business service components. This process of issuing the authentication certificate is managed by the Identity, Management Service (IMS). It is possible to generate a duplicate certificate or forge the certificate and submit it to the web server instead of the original user and identity authentication is bypassed and call to the Business Service Component can be attained. One of the possible implementation solutions for this is to use better communication protocols such as HTTPS, which can't be worked well.

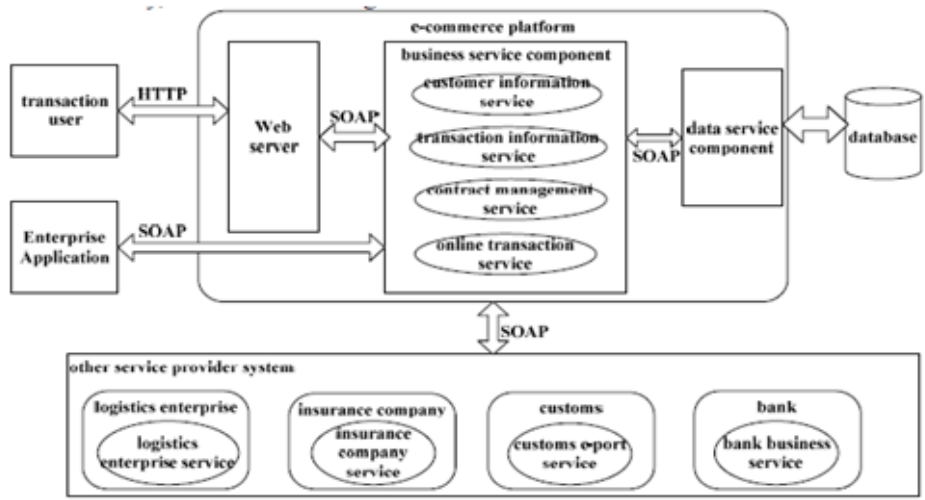

Figure 1: Basic architecture of E-commerce systems

### 3.2 Unsecure Protocol

When petitions are cleared to access the business service components or domain services, identity authentication is required and the request is passed to the corresponding service component through HTTP. Comparatively HTTP is not a secure protocol when it comes to handling MAN–IN-MIDDLE attacks. MAN-IN-MIDDLE attacks are a form of attacks in which attackers gain a connection with receivers and transmitters of messages and attackers relay messages between them giving sender and receiver an illusion that they are communicating directly. The assailant is able to encode the communication between them equally comfortably.

### 3.3 No filters mentioned on the application level

Filters are really important at the application level and the absence of these filters may allow an attacker to send malicious codes through the web application and perform attacks like Cross Site Scripting, Remote/Local File Inclusion etc. which may immediately or indirectly lead to a successful attack.

### 3.4 Unsecure Database

In most of E-commerce models, database is maintained on the same server without going through any security bars. Thither are a number of security risks related to database that are as follows:

• Malware infections into the database may also go to incidents like leakage or disclosure of important information.

• Malfunctioned database may deny service to authorized users and failure of database services can't be foreseen.

### 3.5 Denial of Service Attacks

It brings up to an attempt to establish the system unavailable to its intended users. The existing E-commerce model does not provide any method to prevent the DOS or DDOS attack.

## 4. PROPOSED SECURITY FRAMEWORK

The proposed Security Framework of SOA based E-commerce is shown in Figure 2. The proposed security framework resolves all the above mentioned security threats and risk involved in E-commerce security. E-commerce provides access to various business service components such as Transaction information Services, Online Transaction Services, etc. These business service components can be accessed by the user if they are authorized and authenticated. The authentication of users is handled by the Identity, Management Service (IMS). E-commerce also supports a series of services as well, such as insurance services, banking services, etc. Users can choose any of these business services from the miscellany of the services provided by E-commerce. Users can predict any of these services, by sending access requests to that corresponding service through IMS. The process of IMS is shown in figure 3. For the services to communicate with each other, they need to post a petition to the IMS, which in return generates a certificate of certification. Just when the certificate is successfully approved the services are allowed to communicate with each other.

The approach to business service components can be achieved with two different behaviors which includes land access and remote access. A two tier security is preserved on the web server to insure that the proper validation of authentication is submitted by the user and that the user can't generate a duplicate certificate or forge the certificate. This three level security can be achieved through:

### 4.1 Input Sanitization

To access the business service components, the user submits a request to the web server through a more secure protocol HTTPS. Before sending the service request to the corresponding service component, the user's data are first sanitized and then forwarded to the web server. If the user request consists of some extra qualities such as '#@abc*', then it can clean and handled as 'ABC' with the action of input sanitization rules and the attempt of the user to bypass the authentication can be forestalled. Input sanitization can solve the risks of certificate duplicity, which is one of the important risk factors in E-commerce security.

## 4.2 Rule based Plug-in for additional protection

It allows wrapping an extra layer of protection around the web server. If the sanitization is bypassed by the assailant, it can be clogged by this additional security layer. This layer works as an external security layer, which facilitates to increase protection, such as detection and prevent attacks before the intruder reaches the business service part. To carry out this extra layer of security, for instance, we can use 'Apache mod-security', which is a rule-based plug-in of Apache which allows one to create rules describing abnormal requests to the WWW host. When malicious requests are cleared to access business service components, first they are equated with the rulers that were created if matched, then the requests are denied and the details are logged.

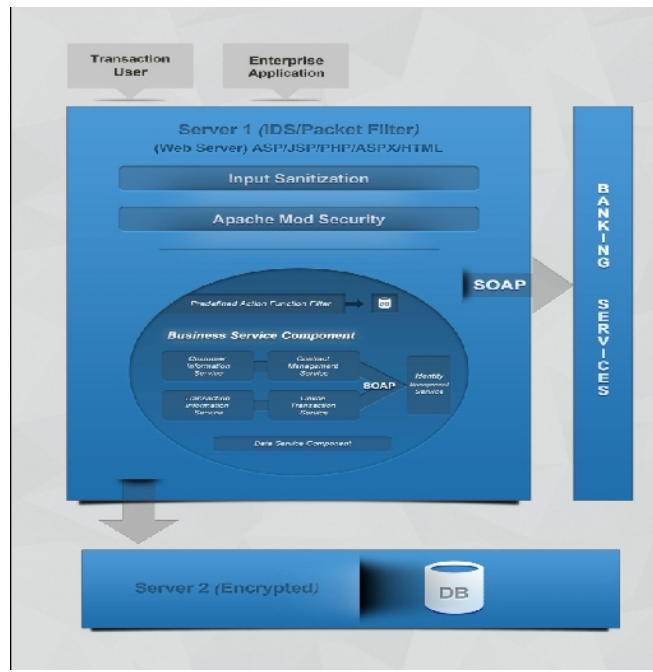

Figure 2: Proposed Security Framework Design

## 4.3 Predefined Action filter

In the proposed security framework filters on the application level is utilized which is responsible for separating out the actions specified on web pages. If somehow intruder bypasses the additional security layer, it enters into predefined action function filters. If the user visits a particular service component and if its corresponding activity is available in the database, then the petition is simply forwarded to the respective service component and the petition will be discharged; and if its action is not available in the database, and so the petition will not be turned over to the service component and an error message will be displayed to the user. Intrusion detection system (IDS) and Intrusion protection system (IPS) is employed in the proposed security framework to provide spare security to business service components and database related. In offering a security framework, business service components and connected database are maintained on different hosts. Server I maintain business service components and network hosts. Server II maintains the databases connected with business service components. The communication between server I and server II is monitored by a firewall. If an attempt is discovered by the Intrusion Detection System (IDS), the firewall will automatically disconnect both servers from one another to assure minimal data loss.

## 5. IMPLEMENTATION BENEFITS OF PROPOSED SECURITY FRAMEWORK

The proposed security framework is implemented and validated on an open source E-commerce website and it proves that it achieves most of the quality attributes. The approach to business service components can be achieved with two different methods, which include field access and remote access. Multiple-layer security is maintained on the web server to insure the proper proof of authentication submitted by the user and that the user can't generate a duplicate certificate or forge the certificate. The main implementation benefits and effects of the proposed plan are as follows:

### 5.1 Database Encryption

The database kept up on a totally different host (Server II), which monitored by IDS/IPS. The database itself encrypted with a unique key which is stashed away in the data service component on a server I. In lawsuit of an attack IDS/IPS, will automatically disconnect all connections between server I and server II to ensure minimum harm. The encryption will provide protection for data being misused by the assailant.

### 5.2 Indirect Access to Enterprise Applications

The Enterprise Applications are not permitted to pass directly with the business services; alternatively, it is linked via the same web host. The application would provide a different user interface and will also ensure the guard of the servers by preventing most attacks.
.
### 5.3 Prevention Against DoS and DDos

The presentation of IDS/IPS on both the servers will prevent most of the Denial of Service attacks as they would monitor and separate out all the incoming information. The IDS/IPS can be easily configured to bar the IP Addresses generating multiple requests within a suitable interval of the fourth dimension.

### 5.4 Post-Exploitation Prevention

This is in case an attacker manages to go around the firewall and somehow run his code.

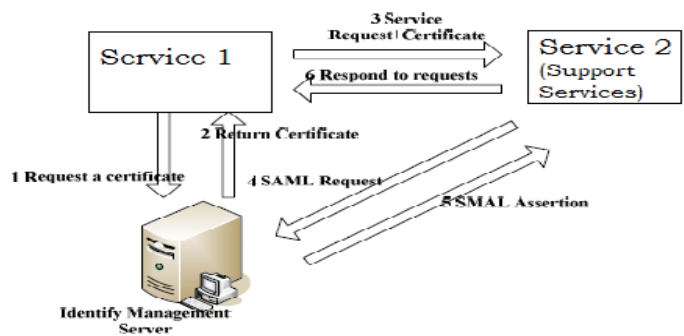

Figure 3: Process of Identity management service

The decisive aim of the security would be to prevent damage from the attempt. Post exploitation prevention can be enabled by editing functions permissions.

## 6. THE FRAMEWORK EVALUATION

The proposed security framework is implemented on an open source E-commerce system through which the proposed framework is examined to discover its strengths and weakness in the light of security threats discussed above. The methodology for evaluating the framework depends upon the character of security threats. The rating also depends on the security requirements and arranging for the endeavor. Some of the character attributes which are already achieved are discussed infra:

### 6.1 Input Sanitization

To access the business service components, the user submits their request to web server through a more secure protocol HTTPS. Before transmitting the service request to the corresponding service component, the user's data are first sanitized and then forwarded to the web host. If a user request consists of some extra qualities such as '#@abc*' as a request, then with the action of input sanitization rules it can clean and handled as 'ABC' and the attempt of user to bypass the authentication can be forestalled. Input sanitization can solve the risks of certificate duplicity, which is one of the important risk factors in E-commerce security. The figure 4.1 shows the infected input submitted by user and figure 4.2 shows the input sanitization for the proposed security plan.

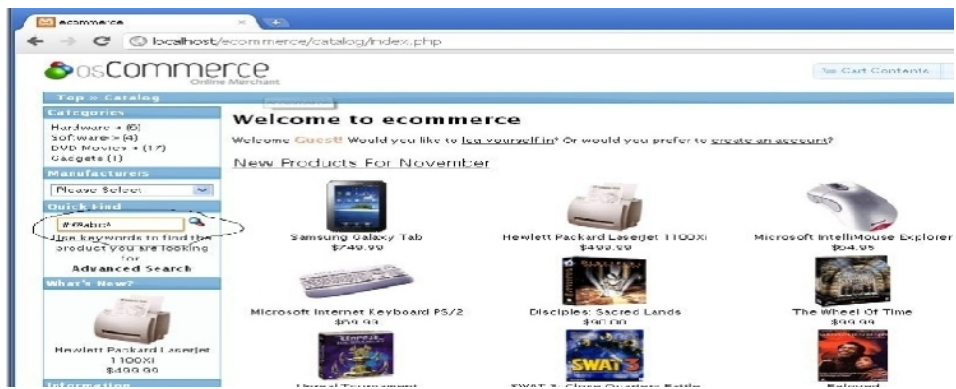

Figure 4.1 users infected input as "#@abc*"

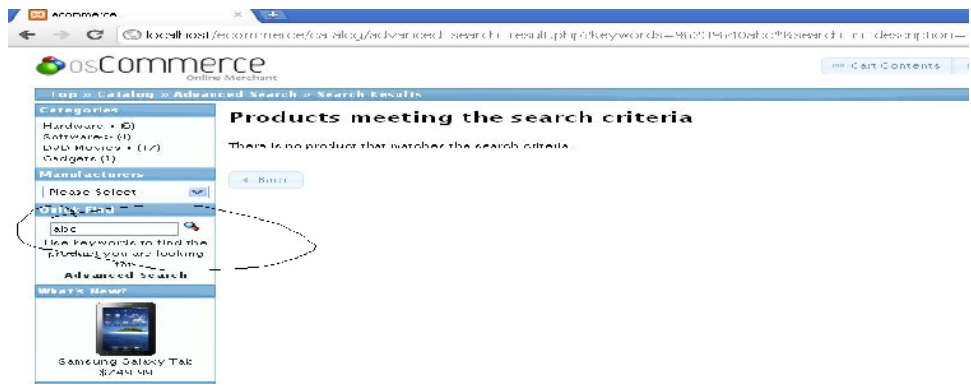

Figure 4.2 the input sanitization for the proposed security design.

## 7. CONCLUSION

This paper presents and analyzes the current security issues for E-commerce system and proposed a security framework for E-commerce system. The proposed framework is based on SOA. The proposed framework provides a secure E-commerce system from the whole of the major computing attacks or threats. The proposed framework is initially implemented on an open source E-commerce system and proves that it reach the reference attributes. For evaluating the proposed security framework all the major computing attacks are tested on. The main contribution and impact of this research is to align the benefits of SOA with E-commerce system, so that organizations can assume the combined benefits of both. The proposed security framework helps enterprise to organize an absolute suite of amalgamated security architecture which protect SOA based E-commerce.